\documentclass[aps,prl,showpacs,twocolumn]{revtex4}
 \ifx\pdftexversion\undefined
\usepackage[dvips]{graphics} \else
\usepackage[pdftex]{graphics} \fi
\usepackage{graphics,epsfig}

\begin{document}
\newtheorem{Theorem}{Theorem}
\title{Arbiter as the Third Man in classical and quantum games}
\author{Jaros\l aw Pykacz$^1$ and Piotr Fr\c{a}ckiewicz}
\affiliation{$^1$Institute of Mathematics, University of
Gda\'{n}sk, 80-952 Gda\'{n}sk, Poland}
\email{pykacz@math.univ.gda.pl, frackiewicz.piotr@wp.pl}
\date{\today}
\begin{abstract}
We study possible influence of not necessarily sincere arbiter on
the course of classical and quantum $2 \times 2$ games and we show
that this influence in the quantum case is much bigger than in the
classical case. Extreme sensitivity of quantum games on initial
states of quantum objects used as carriers of information in a
game shows that a quantum game, contrary to a classical game, is
not defined by a payoff matrix alone but also by an initial state
of objects used to play a game. Therefore, two quantum games
that have the same payoff matrices but begin with different
initial states should be
considered as different games.
\end{abstract}
\maketitle

Exchange of information in the course of a game is often performed
in a non-verbal way. For example players show cards, flip coins,
move figures on a board, etc., which means that they utilize for
this purpose really existing physical objects. Observation that
even some very simple games can change drastically when players
use quantum objects instead of classical ones was the beginning of
the theory of quantum games \cite{Mey99,EWL99,MW00}. Theory of
quantum games is still `in statu nascendi' and till now the
majority of authors studied only the simplest case of static
two-person two-strategy games (see, e.g., \cite{Grab05,Iqb06} for
a review), i.e., games in which each of two players has to choose
one of two possible strategies, and they do it simultaneously not
knowing which strategy is chosen by the other player. Such classical games
are completely defined mathematically by specifying a $2\times 2$
bi-matrix of payoffs that players get when a game is finished, and
can be played even by dumb and illiterate players in the following
way: An arbiter prepares two coins in the `heads-up' state and
passes one of them to each of the players. Each of the players
either does nothing, which means that he rests with his first
strategy, or he flips his coin, which means that he chooses his
second strategy. Then the coins are passed back to the arbiter who
checks the state of both coins (i.e., he makes a measurement!),
and announces the result. This way of playing $2\times 2$ games
can be easily `quantized': it is enough to replace coins by
two-state quantum objects (qubits), while leaving all the rest of
the scheme unchanged. Actually, this way of `quantizing' the
classical Battle of The Sexes game was proposed by Marinatto and
Weber in \cite{MW00} and it is, according to our opinion, the most
natural way of `quantizing' classical games (the scheme proposed
by Eisert, Wilkens, and Lewenstein in \cite{EWL99} allows in the
quantum case strategies that are not allowed in the classical
case, therefore the original game is in fact replaced by another
one \cite{EP02}).

However, when a $2\times 2$ quantum game is played according to
the Marinatto-Weber scheme the role of the arbiter is no passive
anymore since it is up to him what initial state of two
coins/qubits he passes to the players. Marinatto and Weber in
\cite{MW00}, as well as other authors writing papers on quantum
games, did their best in order to `save' the original classical
game in their quantum scheme: usually the original classical game
was recovered from the quantum one when the initial state of two
qubits prepared by the arbiter was not entangled.

In this Letter we go beyond this constraint and allow the arbiter to
be the Third Man in a game. This means that he can cheat the
players by sending them coins/qubits in the other initial state
than they expect. Since the players do not know the actual state
of coins/qubits they get, they choose their strategies according
to their \textit{belief}, not according to the \textit{actual
state} of coins/qubits that are sent to them. In order to be
illustrative, we confine our considerations to symmetric 2x2 games
in which no payoffs are identical since any such game falls into
one of the three categories \cite{Wei95}: I (Prisoner's Dilemma
Game), II (Coordination Game), or III (Hawk-Dove Game). We study
what changes of such a game can be induced by a `classical
cheating arbiter', i.e., by the arbiter that can pass to the
players two coins in the initial state (H,T), or (T,H), or (T,T),
instead of the presumed initial state (H,H). We show that he can
only change a symmetric game into a non-symmetric one but he
cannot change a category of a game if it remains symmetric. Then,
we show that for a `quantum cheating arbiter' that passes to the
players two qubits that are not in the presumed initial state
$|00>$ but are in another state that is allowed by quantum
mechanics the variety of possible changes is much bigger which
allows him to exert his influence on the course of a game in many
ways that are impossible for his `classical' colleague.

\textit{Symmetric $2\times 2$ games.}
---Any symmetric two-player game is fully characterized by the
pure-strategy payoff matrix to one of the players
\begin{equation}
A= \left[\begin{array}{cccc}
a_{11} & a_{12} & \cdots & a_{1n} \\
a_{21} & a_{22} & \cdots & a_{2n} \\
\vdots & \vdots & \ddots & \vdots \\
a_{n1} & a_{n2} & \cdots & a_{nn}
\end{array}\right]
\end{equation}
since the second player's payoff matrix is the transpose of the
matrix $A$: $B=A^T$. A value of an element $a_{ij}$ of the matrix
$A$ is the payoff to the player A when he plays an i-th pure
strategy and his opponent plays a j-th strategy.

In the case when each player has only two pure strategies and no
payoffs are identical there are only three generic categories of
symmetric two-player games. They can be distinguished according to
signs of differences $\alpha =a_{11}-a_{21}$ and $\beta
=a_{22}-a_{12}$, and are as follows (see, e.g., \cite{Wei95}):

\noindent \textbf{Category I} (Prisoner's Dilemma Game): $\alpha
\beta <0$.

\noindent \textbf{Category II} (Coordination Game): $\alpha ,
\beta >0$.

\noindent \textbf{Category III} (Hawk-Dove Game): $\alpha , \beta
<0$.

It occurs that within each category various games differ only with
respect to specific values of payoffs but possess the same
dominance relations, Nash equilibria, etc., i.e., all games that
belong to the same category should be played in the same way.

\textit{Arbiter as the `Third Man' in classical games.}---
Let us now consider symmetric $2\times 2$ games played in a previously
mentioned way that allows an arbiter to be the
Third Man in a game. Information between the arbiter and players
is conveyed by a pair of classical two-state objects, e.g., two
coins. In the beginning the arbiter passes to each of the players
one coin in the `heads-up' state that symbolizes the first
strategy. Each of the players can either return to the arbiter his
coin without changing its state, which means that he remains with
the first strategy, or he can flip his coin informing in this way
the arbiter that he chooses the second strategy. Of course this
way of exchanging information does not make the game different
from the same game played in any other way, although it can be
played now even by dumb and illiterate players. However, let us
assume that the players are blind and that the arbiter is not
sincere and can cheat the players passing them in the beginning of
the game two coins that are not in a (`heads-up',`heads-up') =
(H,H) state but are in another possible state: (H,T), (T,H), or
(T,T). In such a situation blind players who do not expect that
the arbiter can cheat them will choose their strategies according
to their \textit{belief}, not according to the \textit{actual
state} of coins they get and this will surely lead to a lot of
confusion. We think, however, that the considered problem is not
purely artificial and that such situations can be encountered also
in the real life, e.g., at the stock exchange where players
sometimes do not know what game they are actually playing.

Although the sketched situation might give rise to numerous
interesting problems, in this paper we shall concentrate on the
following one: what changes of categories of a symmetric $2\times
2$ game in which no payoffs are identical can be made by a
`cheating arbiter'. The answer for a classical arbiter is
contained in the following

\textit{Theorem 1.}---A cheating classical arbiter cannot change a
category of symmetric $2\times 2$ game in which no payoffs are
identical. He can only change such game into a non-symmetric game.

\textit{Proof.}---Let
\begin{equation}
(A,B)= \left[\begin{array}{cc}
(x,x) & (y,w) \\
(w,y) & (z,z) \\
\end{array}\right]
\label{bimatrix}
\end{equation}
where all numbers $x,y,w,z$ are different, be a bi-matrix of a
symmetric $2\times 2$ game. If the arbiter cheats players by
sending them two coins that are in the (H,T) state instead of the
expected (H,H) state, the second player chooses the second
strategy thinking that he has chosen the first strategy and vice
versa, so now the bi-matrix of the game becomes
\begin{equation}
(A',B')= \left[\begin{array}{cc}
(y,w) & (x,x) \\
(z,z) & (w,y) \\
\end{array}\right]
\end{equation}
i.e., it is the original bi-matrix of payoffs with permuted
columns. However, since
\begin{equation}
(A')^T= \left[\begin{array}{cc}
y & z \\
x & w \\
\end{array}\right]  \neq B'=  \left[\begin{array}{cc}
w & x \\
z & y \\
\end{array}\right],
\end{equation}
and all payoffs are different, a game characterized by the
bi-matrix $(A',B')$ is not symmetric. The same arguments apply
when the arbiter sends to the players coins in the (T,H)  state
instead of  the expected state (H,H).

Finally, if the arbiter sends to the players coins in the state
(T,T) instead of (H,H), then both players play the `opposite'
strategies than they expect, and the bi-matrix of payoffs becomes
\begin{equation}
(A'',B'')= \left[\begin{array}{cc}
(z,z) & (w,y) \\
(y,w) & (x,x) \\
\end{array}\right].
\end{equation}
A game characterized by such bi-matrix of payoffs is symmetric,
but if for the original game $\alpha =x-w$, $\beta =z-y$, then for
the game characterized by the bi-matrix $(A'',B'')$  $\alpha
''=z-y=\beta $ and $\beta ''=x-w=\alpha $. Therefore, a new game
remains in the same category as the original one, which finishes
the proof.

\textit{Quantum games played according to the Marinatto-Weber scheme.}---
In one of the first papers on quantum games Marinatto and Weber
\cite{MW00} `quantized' the classical Battle of The Sexes game in
a way that essentially boils down to replacing classical carriers
of information, e.g., coins mentioned in the previous section, by
two-state quantum objects (qubits) while leaving all the rest of
the procedure unchanged. According to their idea a $2\times 2$
quantum game should be played as follows: An arbiter prepares a
pair of two-state quantum objects in a specific state, e.g., two
spins in a spin-up state, and sends one of these objects to each
of the players. Each player either returns his object to the
arbiter unchanged, i.e., applies to it an identity operation, or
applies to the object an operation that changes its state into the
opposite state, e.g., in the case of spins he flips the spin. Then
the arbiter measures the state of both objects he got back from
the players and announces the result taking into account the
payoff matrix of the game. It is obvious that the only difference
between playing classical $2\times 2$ games in a way described in
the previous section and playing quantum $2\times 2$ games
according to the Marinatto-Weber scheme is such that in the former
information is carried by classical and in the latter by quantum
objects.

However, while a pair of classical two-state objects can be only
in one of four possible pure states since its space of pure states
is a Cartesian product of two two-element sets, a pair of quantum
objects possesses infinity of pure states, in particular various
superpositions of `basic' states that represent pure classical
strategies, and even `less classical' entangled states, which are
a main source of differences between classical and quantum games.
In particular, when a pair of quantum objects used to play a game
is in an entangled state, the well-known EPR correlations
introduce a kind of `unconscious communication' between players
that does not exist in the classical case and which allowed
Marinatto and Weber to find new solutions of the Battle of The
Sexes game.

The way of playing quantum games proposed by Marinatto and Weber
is not the only one proposed in the literature, but according to
our opinion a quantum game played in this way differs the least
from its classical prototype. Finally, the players even do not
have to know whether carriers of information used in their
game are classical or quantum: they might be informed only that
they either should do nothing, or press a button causing in this
way a `flip' of a state of an object sent by the arbiter. Although
the majority of authors writing on quantum games seems to follow
Eisert-Wilkens-Lewenstein \cite{EWL99} and admit that players may
use as an allowed  strategy any unitary operation, we rather agree
with van Enk and Pike \cite{EP02} that changing the set of allowed
strategies means changing the game itself, and therefore we prefer
the scheme proposed by Marinatto and Weber.

\textit{Arbiter as the `Third Man' in quantum games.}---
In the case of quantum games played according to the
Marinatto-Weber scheme a cheating arbiter has more possibilities
of exerting his influence on a game since he can send to the
players a pair of qubits that is not, as the players believe, in
the state $|00>$ but is in any state of the form
\begin{equation}
|\psi_{in}> = a|00> + b|01> + c|10> + d|11>
\label{initial state}
\end{equation}
where complex coefficients satisfy the normalization requirement
$|a|^2 + |b|^2 +|c|^2 + |d|^2 = 1$.  Let us stress that our aim is
exactly opposite to the aim of all authors writing up to now
papers on quantum games: while they did their best in order to
keep the original classical game as a `special case' of a quantum
game, we want to study  what \textit{changes} of a game can be
made by a `cheating quantum arbiter'. Since we would like to stay,
however, within the category of symmetric games, we begin with the
following

\textit{Theorem 2.}---If in the initial state (\ref{initial state})
prepared by an arbiter
$|b|=|c|$, then a symmetric $2\times 2$ quantum game in which no
payoffs are identical remains symmetric after the intervention of
the arbiter. Moreover, if the payoff matrix (\ref{bimatrix}) of a game is such
that $x \neq z$ or $y \neq w$, then the condition $|b|=|c|$ is
also necessary for preserving symmetry of a game.

\textit{Proof.}---Following calculations of Marinatto and
Weber \cite{MW00} one can check that if the bi-matrix of the
original game is (2),  then the initial state $|\psi_{in}>$ (6)
yields the following payoff matrices for the player A
\begin{equation}
\left[\begin{array}{cc} \! x|a|^2 \! + \! y|b|^2 \! + \! w|c|^2 \!
+ \! z|d|^2 & x|b|^2 \! + \! y|a|^2 \! + \! w|d|^2 \! + \! z|c|^2
\! \\ \! x|c|^2 \! + \! y|d|^2 \! + \! w|a|^2 \! + \! z|b|^2 &
x|d|^2 \! + \! y|c|^2 \! + \! w|b|^2 \! + \! z|a|^2 \!  \\
\end{array}\right]
\label{final bimatrix}
\end{equation}
and for the player B
\begin{equation}
\left[\begin{array}{cc} \! x|a|^2 \! + w|b|^2 \! + y|c|^2 \! + \!
z|d|^2 & x|b|^2 \! + \!
w|a|^2 \! + \! y|d|^2 \! + \! z|c|^2 \\
x|c|^2 \! + \! w|d|^2 \! + \! y|a|^2 \! + \! z|b|^2 & x|d|^2 \!
+ \! w|c|^2 \! + \! y|b|^2 \! + \! z|a|^2 \! \\
\end{array}\right].
\end{equation}
Thus, the requirement $A^T = B$ is equivalent to constraints
expressed by the following system of equations:
\begin{equation} \begin{array}{l}
(|b|^2 - |c|^2)(x-z) = 0\\ (|b|^2 - |c|^2)(y-w) = 0\\
\end{array}
\label{equations}
\end{equation}
We see that when $|b| =|c|$ these constraints are fulfilled whatever
are values of payoffs.

On the other hand, if $x \neq z$ or $y \neq w$, then the condition
$|b| = |c|$ is a necessary condition to make equations (\ref{equations})
satisfied, which finishes the proof.

From this point on we shall assume that an intervention of the
arbiter into a game is such that the game remains symmetric.  In
such case we get the following

\textit{Theorem 3.}---A cheating quantum arbiter can, retaining symmetry of the
original $2 \times 2$ game in which no payoffs are identical,
change its payoff bi-matrix without changing its category, as well
as, in almost all cases, change its category into any other one.

\textit{Proof}---Of course only the initial state
$|\psi_{in}> = a|00>$ with $|a|=1$ causes no changes in the
original bi-matrix (2) of a game.

Let us denote $\alpha _0 = x-w$ and $\beta _0 = z-y$
characteristic numbers of the initial game. Then, assuming that
$|b|=|c|$, one can calculate from the matrix (7) of the final game
its characteristic numbers
\begin{equation}
\alpha = \alpha _0 (|a|^2 - |b|^2) + \beta _0 (|d|^2-|b|^2)
\end{equation}
\begin{equation}
\beta = \alpha _0 (|d|^2-|b|^2) + \beta _0 (|a|^2 - |b|^2).
\end{equation}

Let us assume that the initial game belongs to category I, i.e.,
that $\alpha _0 \beta _0 < 0$. Then if we put $|a|^2 < |b|^2 =
|c|^2 < |d|^2 $ in the initial state (6), we get $\alpha\beta < 0$
so the final game also belongs to category I. If the initial game
belongs to category II or III, one can easily check that the
condition $|b|^2 < \min(|a|^2,|d|^2)$ is sufficient to keep
category of the game unchanged.

Since a `cheating arbiter', while preparing the initial state
(\ref{initial state}) of a game is constrained only by the normalization
condition and, if he wants to retain symmetry of a game, by the condition
$|b|=|c|$, it is obvious that in general he can choose
coefficients $a,b,c,$ and $d$ in the initial state (\ref{initial state}) in such a
way that category of a game changes into any other one according
to his will. For example, if he prepares the initial state (\ref{initial state}) so
that $|b|^2 > \max(|a|^2,|d|^2)$, then he changes category of a
game from II to III and vice versa. It can be also checked that if
$\alpha _0 \neq \pm \beta _0$ and the initial state (\ref{initial state}) is such
that $|a|^2 \neq |d|^2$ and $|b|^2 = 1/4$, then a game changes its
category from II or III to I, and the opposite change is forced by
putting in (6) $|a|^2 = |d|^2 \neq |b|^2$.

However, not in every case any conceivable change is possible. One
can check, for example, that if $\alpha _0 = \beta _0$ and the
original game belongs to category II (resp. III), then it can be
changed into a game that belongs to category III (resp. II), but
not into a game that belongs to category I. Even worse situation
occurs when the original game belongs to category I and $\alpha _0
= -\beta _0$: In this case an arbiter cannot change category of a
game.

However, such cases are exceptional and in general a `cheating
arbiter' can change category of a symmetric $2 \times 2$ quantum
game into any other one according to his will, so the proof is finished.

\textit{Concluding remarks.}---Our case studies of the influence
of a `cheating arbiter' on symmetric $2 \times 2$ games confirm
the general belief that the range of possibilities in the
`quantum' case is always much wider than in the `classical' case.
Moreover, extreme sensitivity of $2 \times 2$ quantum games on the
specific form of the initial state of a pair of qubits used to
play a game (called in \cite{MW00} `a strategy') suggests changing
a way of looking at quantum games. Classical static games are
completely defined by their matrices of payoffs and this way of
looking at quantum games was up to now adopted also in the quantum
domain. Of course it was noticed already in the very first papers
on quantum games that players who play a quantum game should base
their decissions not only on the analysis of a payoff matrix of a
game but also on the initial state of a pair of qubits used as
carriers of information in a game, however, it was never
considered who and according to what rules prepares this initial
state. Since two quantum games that have the same payoff matrices
but begin with different initial states of quantum objects used as
carriers of information usually force the players to choose
different strategies, in our opinion such two quantum games are in
fact different games. Therefore, we argue that static quantum
games, contrary to their classical prototypes are not fully
defined by specifying their payoff matrices alone and defining
them requires also defining the specific initial state of a set of
quantum objects used to play a quantum game.

Finally, let us note that although players that play a classical
static game can exchange information about chosen strategies using for
this purpose various physical objects, usually they can do it also
verbally or in writing. This way of playing quantum games, except
of a possibility of playing $2 \times 2$ quantum games according
to Marinatto-Weber scheme with the use of macroscopic objects
\cite{AHPPDM07} is still, in general, not known.

This work has been supported by the University of Gda\'{n}sk research
grants BW/5100-5-0203-6 and BW/5100-5-0040-7.

\end{document}